\documentclass[twocolumn]{aastex631}
\usepackage{graphicx}
\usepackage{float}
\usepackage{amsmath}
\usepackage{amssymb}
\usepackage{subfigure}
\usepackage{cases}
\usepackage{mathrsfs}
\usepackage[flushleft]{threeparttable}
\usepackage{amssymb}
\usepackage{pifont}
\usepackage{epstopdf}
\usepackage{xcolor}
\usepackage[all]{hypcap}
\usepackage{mwe,tikz}
\usepackage{bm}
\usepackage{multirow}
\usepackage{longtable}
\usepackage{xspace}
\usepackage{rotating}
\usepackage{CJK}
\usepackage{url}
\usepackage{upgreek}
\usepackage{booktabs}
\usepackage{textcomp}
\usepackage{makecell}
\usepackage{enumitem}
\usepackage{csquotes}
\usepackage{soul}
\usepackage{tabularx}

\newcommand{\e}{\text{e}}

\newcommand{\p}{\partial}

\begin{document}

\title{Triggering the Untriggered: The First Einstein Probe-Detected Gamma-Ray Burst 240219A and Its Implications}

\correspondingauthor{Bin-Bin Zhang, Chen Zhang}
\email{bbzhang@nju.edu.cn; chzhang@bao.ac.cn}

\author[0000-0002-5596-5059]{Yi-Han Iris Yin}
\affiliation{School of Astronomy and Space Science, Nanjing University, Nanjing 210093, China}
\affiliation{Key Laboratory of Modern Astronomy and Astrophysics (Nanjing University), Ministry of Education, China}

\author[0000-0003-4111-5958]{Bin-Bin Zhang}
\affiliation{School of Astronomy and Space Science, Nanjing University, Nanjing 210093, China}
\affiliation{Key Laboratory of Modern Astronomy and Astrophysics (Nanjing University), Ministry of Education, China}
\affiliation{Purple Mountain Observatory, Chinese Academy of Sciences, Nanjing 210023, China}

\author[0000-0002-5485-5042]{Jun Yang}
\affiliation{School of Astronomy and Space Science, Nanjing University, Nanjing 210093, China}
\affiliation{Key Laboratory of Modern Astronomy and Astrophysics (Nanjing University), Ministry of Education, China}

\author[0000-0002-9615-1481]{Hui Sun}
\affiliation{National Astronomical Observatories, Chinese Academy of Sciences, Beijing 100101, China}

\author{Chen Zhang}
\affiliation{National Astronomical Observatories, Chinese Academy of Sciences, Beijing 100101, China}

\author{Yi-Xuan Shao}
\affiliation{School of Astronomy and Space Science, Nanjing University, Nanjing 210093, China}
\affiliation{Key Laboratory of Modern Astronomy and Astrophysics (Nanjing University), Ministry of Education, China}

\author{You-Dong Hu}
\affiliation{Instituto de Astrof\'isica de Andaluc\'ia (IAA-CSIC), Glorieta de la Astronom\'ia s/n, E-18008, Granada, Spain}

\author{Zi-Pei Zhu}
\affiliation{National Astronomical Observatories, Chinese Academy of Sciences, Beijing 100101, China}

\author{Dong Xu}
\affiliation{National Astronomical Observatories, Chinese Academy of Sciences, Beijing 100101, China}

\author{Li An}
\affiliation{School of Physics and Astronomy, Beijing Normal University, Beijing 100875, China}
\affiliation{Institute for Frontier in Astronomy and Astrophysics, Beijing Normal University, Beijing 102206, China}

\author{He Gao}
\affiliation{School of Physics and Astronomy, Beijing Normal University, Beijing 100875, China}
\affiliation{Institute for Frontier in Astronomy and Astrophysics, Beijing Normal University, Beijing 102206, China}

\author{Xue-Feng Wu}
\affiliation{Purple Mountain Observatory, Chinese Academy of Sciences, Nanjing 210023, China}

\author[0000-0002-9725-2524]{Bing Zhang}
\affiliation{Nevada Center for Astrophysics, University of Nevada Las Vegas, NV 89154, USA}
\affiliation{Department of Physics and Astronomy, University of Nevada Las Vegas, NV 89154, USA}

\author{Alberto Javier Castro-Tirado}
\affiliation{Instituto de Astrof\'isica de Andaluc\'ia (IAA-CSIC), Glorieta de la Astronom\'ia s/n, E-18008, Granada, Spain}
\affiliation{Departamento de Ingenier\'ia de Sistemas y Autom\'atica, Escuela de Ingenier\'ias, Universidad de M\'alaga, C\/. Dr. Ortiz Ramos s\/n, E-29071, M\'alaga, Spain}

\author{Shashi B. Pandey}
\affiliation{Aryabhatta Research Institute of Observational Sciences (ARIES), Manora Peak, Nainital-263002, India}

\author{Arne Rau}
\affiliation{Max-Planck-Institut für Extraterrestrische Physik, Gießenbachstraße, D-85748 Garching, Germany}

\author{Weihua Lei}
\affiliation{Department of Astronomy, School of Physics, Huazhong University of Science and Technology, Wuhan, Hubei 430074, China}

\author{Wei Xie}
\affiliation{Department of Astronomy, School of Physics and Electronic Science, Guizhou Normal University, Guiyang, 550001, China}

\author{Giancarlo Ghirlanda}
\affiliation{INAF - Osservatorio Astronomico di Brera, Via E Bianchi 46, I-23807 Merate (LC), Italy}
\affiliation{INFN - Sezione di Milano-Bicocca, piazza della Scienza 3, I-20126 Milano (MI), Italy}

\author{Luigi Piro}
\affiliation{INAF - Istituto di Astrofisica e Planetologia Spaziali, via Fosso del Cavaliere 100, 00133 Rome, Italy}

\author{Paul O'Brien}
\affiliation{School of Physics and Astronomy, University of Leicester, University Road, Leicester, LE1 7RH, UK}

\author{Eleonora Troja}
\affiliation{Department of Physics, University of Rome ``Tor Vergata'', via della Ricerca Scientifica 1, I-00133 Rome, Italy}
\affiliation{INAF - via Parco del Mellini, 00100 Rome, Italy}

\author{Peter Jonker}
\affiliation{Department of Astrophysics/IMAPP, Radboud University Nijmegen, Nijmegen, 6500 GL, The Netherlands}

\author{Yun-Wei Yu}
\affiliation{Institute of Astrophysics, Central China Normal University, Wuhan 430079, China}

\author{Jie An}
\affiliation{National Astronomical Observatories, Chinese Academy of Sciences, Beijing 100101, China}

\author{Run-Chao Chen}
\affiliation{School of Astronomy and Space Science, Nanjing University, Nanjing 210093, China}
\affiliation{Key Laboratory of Modern Astronomy and Astrophysics (Nanjing University), Ministry of Education, China}

\author{Yi-Jing Chen}
\affiliation{School of Astronomy and Space Science, Nanjing University, Nanjing 210093, China}

\author{Xiao-Fei Dong}
\affiliation{School of Astronomy and Space Science, Nanjing University, Nanjing 210093, China}
\affiliation{Key Laboratory of Modern Astronomy and Astrophysics (Nanjing University), Ministry of Education, China}

\author{Rob Eyles-Ferris}
\affiliation{School of Physics and Astronomy, University of Leicester, University Road, Leicester, LE1 7RH, UK}

\author{Zhou Fan}
\affiliation{National Astronomical Observatories, Chinese Academy of Sciences, Beijing 100101, China}

\author{Shao-Yu Fu}
\affiliation{National Astronomical Observatories, Chinese Academy of Sciences, Beijing 100101, China}

\author{Johan P.U. Fynbo}
\affiliation{Cosmic Dawn Center (DAWN), Copenhagen 2200, Denmark}
\affiliation{Niels Bohr Institute, University of Copenhagen, Copenhagen 2200, Denmark}

\author{Xing Gao}
\affiliation{Xinjiang Astronomical Observatory, Chinese Academy of Sciences, Urumqi, Xinjiang 830011, China}

\author{Yong-Feng Huang}
\affiliation{School of Astronomy and Space Science, Nanjing University, Nanjing 210093, China}
\affiliation{Key Laboratory of Modern Astronomy and Astrophysics (Nanjing University), Ministry of Education, China}

\author{Shuai-Qing Jiang}
\affiliation{National Astronomical Observatories, Chinese Academy of Sciences, Beijing 100101, China}

\author{Ya-Hui Jiang}
\affiliation{School of Astronomy and Space Science, Nanjing University, Nanjing 210093, China}

\author{Yashaswi Julakanti}
\affiliation{School of Physics and Astronomy, University of Leicester, University Road, Leicester, LE1 7RH, UK}

\author{Erik Kuulkers}
\affiliation{European Space Agencay, ESTEC, Keplerlaan 1, 2201 AZ Noordwijk, The Netherlands}

\author{Qing-Hui Lao}
\affiliation{School of Astronomy and Space Science, Nanjing University, Nanjing 210093, China}

\author{Dongyue Li}
\affiliation{National Astronomical Observatories, Chinese Academy of Sciences, Beijing 100101, China}

\author{Zhi-Xing Ling}
\affiliation{National Astronomical Observatories, Chinese Academy of Sciences, Beijing 100101, China}

\author{Xing Liu}
\affiliation{National Astronomical Observatories, Chinese Academy of Sciences, Beijing 100101, China}

\author{Yuan Liu}
\affiliation{National Astronomical Observatories, Chinese Academy of Sciences, Beijing 100101, China}

\author{Jia-Yu Mou}
\affiliation{School of Astronomy and Space Science, Nanjing University, Nanjing 210093, China}


\author{Xin Pan}
\affiliation{National Astronomical Observatories, Chinese Academy of Sciences, Beijing 100101, China}

\author{Varun}
\affiliation{School of Astronomy and Space Science, Nanjing University, Nanjing 210093, China}

\author{Daming Wei}
\affiliation{Purple Mountain Observatory, Chinese Academy of Sciences, Nanjing 210023, China}

\author{Qinyu Wu}
\affiliation{National Astronomical Observatories, Chinese Academy of Sciences, Beijing 100101, China}

\author{Muskan Yadav}
\affiliation{Department of Physics, University of Rome ``Tor Vergata'', via della Ricerca Scientifica 1, I-00133 Rome, Italy}

\author[0000-0003-0691-6688]{Yu-Han Yang}
\affiliation{Department of Physics, University of Rome ``Tor Vergata'', via della Ricerca Scientifica 1, I-00133 Rome, Italy}

\author{Weimin Yuan}
\affiliation{National Astronomical Observatories, Chinese Academy of Sciences, Beijing 100101, China}

\author{Shuang-Nan Zhang}
\affiliation{Key Laboratory of Particle Astrophysics, Institute of High Energy Physics, Chinese Academy of Sciences, Beijing 100049, China}

\begin{abstract}
The Einstein Probe (EP) achieved its first detection and localization of a bright X-ray flare, EP240219a, on 2024 February 19, during its commissioning phase. Subsequent targeted searches triggered by the EP240219a alert identified a faint, untriggered gamma-ray burst (GRB) in the archived data of Fermi Gamma-ray Burst Monitor (GBM), Swift Burst Alert Telescope (BAT), and Insight-HXMT/HE. The EP Wide-field X-ray Telescope (WXT) light curve reveals a long duration of approximately 160 s with a slow decay, whereas the Fermi/GBM light curve shows a total duration of approximately 70 s. The peak in the Fermi/GBM light curve occurs slightly later with respect to the peak seen in the EP/WXT light curve. Our spectral analysis shows that a single cutoff power-law (PL) model effectively describes the joint EP/WXT--Fermi/GBM spectra in general, indicating coherent broad emission typical of GRBs. The model yielded a photon index of $\sim -1.70 \pm 0.05$ and a peak energy of $\sim 257 \pm 134$ keV. After detection of GRB 240219A, long-term observations identified several candidates in optical and radio wavelengths, none of which was confirmed as the afterglow counterpart during subsequent optical and near-infrared follow-ups. The analysis of GRB 240219A classifies it as an X-ray rich GRB (XRR) with a high peak energy, presenting both challenges and opportunities for studying the physical origins of X-ray flashes, XRRs, and classical GRBs. Furthermore, linking the cutoff PL component to nonthermal synchrotron radiation suggests that the burst is driven by a Poynting flux-dominated outflow.

\end{abstract}
\section{Introduction}
\label{sec:intro}
The Einstein Probe (EP), launched on 2024 January 9, is dedicated to monitoring the sky in the soft X-ray band \citep{2022hxga.book...86Y}. Equipped with two scientific instruments, the Wide-field X-ray Telescope (WXT) and the Follow-up X-ray Telescope (FXT), EP offers a large instantaneous field of view for detecting rapid transients, along with a considerable effective area crucial for follow-up observations and precise localization. Due to its exceptional sensitivity in detecting rapid transients, EP holds significant promise for identifying faint, high-redshift, or newly discovered high-energy transients, especially gamma-ray bursts (GRBs). 

On 2024 February 19 at 06:21:42 UT (referred to as $T_{\rm 0}$), EP/WXT detected and located a bright X-ray flare named EP240219a \citep{2024ATel16463....1Z, 2024ATel16472....1Z} during its commissioning phase at R.A. = 80.$^{\circ}$031 and decl. = 25.$^{\circ}$533 (J2000) with an uncertainty of 2.$'$3 (Figure \ref{fig:obs}). The rapid follow-up instrument on board, EP/FXT, had not yet been turned on when EP240219a was detected. EP/FXT was officially switched on as scheduled during the EP in-orbit testing on 2024 February 28. The source exhibits an overall profile of a fast-rise and exponential-decay shape and lasts approximately 160 s (see Figure \ref{fig:lc}). Fitted by an absorbed power law (PL), the time-averaged spectrum can be characterized with an absorption $N_{\rm{H}}$ of $1.03_{-0.21}^{+0.39} \times\ 10^{22}\ \rm{cm^{-2}}$ and a photon index of $-1.94_{-0.66}^{+0.36}$. The calculated time-averaged unabsorbed fluence in the 0.5 - 4.0 keV range is $7.85_{-1.51}^{+4.06}\times10^{-7}\,\rm{erg\,cm^{-2}}$ (see Table \ref{tab:summary}). 

Following the EP alert, a faint, untriggered gamma-ray transient that occurred at the same time as $T_{\rm 0}$ was discovered offline in the archived data of Fermi Gamma-Ray Burst Monitor \citep[GBM;][]{2024ATel16473....1Z, 2024GCN.35773....1Z} and confirmed by the GBM team's targeted search \citep{2024GCN.35776....1F}, Swift Burst Alert Telescope \citep[BAT;][]{2024GCN.35784....1D} and Insight-HXMT/HE \citep{2024GCN.35785....1C}. The chance coincidence is $\sim 7.85\times 10^{-7}$ given the Swift/BAT and EP/WXT spatial and temporal information \citep{2024GCN.35784....1D, 2024ATel16463....1Z, 2024ATel16472....1Z}.
The gamma-ray transient exhibits a total duration of approximately 70 s and shared a consistent location with EP240219a, leading to its reclassification as GRB 240219A. The time-integrated and time-resolved spectral analysis indicate a peak energy of 127 keV. Applying the fluence ratio criterion $0.72 < r = S(25-50\/ \rm{keV})$ / $S(50-100\/ \rm{keV}) \leq 1.32$ established by \citet{Sakamoto_2008}, this event is classified as an X-ray rich gamma-ray burst (XRR). Follow-up observations identified one potential optical source and seven potential radio sources \citep{2024GCN.35788....1H}. The optical candidate was subsequently ruled out by follow-up observations. Six of the seven radio sources are associated with near-infrared or optical counterparts (Figure \ref{fig:obs}), but none were verified to be linked to the GRB \citep{2024GCN.35808....1K, 2024GCN.35865....1S}.

In this Letter, we report on the detection of EP240219a/GRB 240219A and its follow-up observations, including a multiwavelength analysis and an exploration of its implications. This Letter is structured as follows. Section \ref{sec:data} provides the data reduction and analysis and presents the results. Section \ref{sec:obs} details the long-term follow-up observations and possible afterglow candidates. In Section \ref{sec:nature}, we explore the implications of the burst's classification and spectral components. Finally, Section \ref{sec:sum} summarizes the characteristics and findings derived from this first GRB observed by EP.

\begin{table}
\centering
\begin{threeparttable}
\caption{Summary of the observed properties of GRB 240219A. All errors represent the 1 $\sigma$ uncertainties.}
\label{tab:summary}
\begin{tabular}{ll}
\hline
\hline
Observed Properties & GRB 240219A \\
\hline
$T_{\rm 90, \gamma}$ ($\rm s$) & $54.8_{-4.2}^{+6.2}$ \\
$T_{\rm 90, X}$ ($\rm s$) & $129.3_{-4.4}^{+7.7}$ \\
MVT ($\rm s$) & $\sim 2.19$ \\
Spectral index $\alpha$ & $-1.70_{-0.05}^{+0.05}$ \\
{Peak energy} ($\rm keV$) & $257_{-134}^{+132}$ \\
Gamma-ray fluence ($\rm erg\,cm^{-2}$) & $2.44_{-0.34}^{+0.69}\times10^{-6}$ \\
X-ray fluence\tnote{a} ($\rm erg\,cm^{-2}$) & $7.85_{-1.51}^{+4.06}\times10^{-7}$ \\
Fluence ratio\tnote{b} & $0.85_{-0.10}^{+0.15}$ \\
\hline
\hline
\end{tabular}
\begin{tablenotes}
\item [a] The X-ray fluence is calculated in the energy range of 0.5--4.0 keV.
\item [b] The fluence ratio is obtained between the energy ranges of 25--50 keV and 50--100 keV using the joint spectral fitting results for 0--70 s (Table \ref{tab:spec_fit}).
\end{tablenotes}
\end{threeparttable}
\end{table}

\section{Data Reduction and Analysis}
\label{sec:data}
\begin{figure*}
 \centering
 \includegraphics[width = 0.9\textwidth]{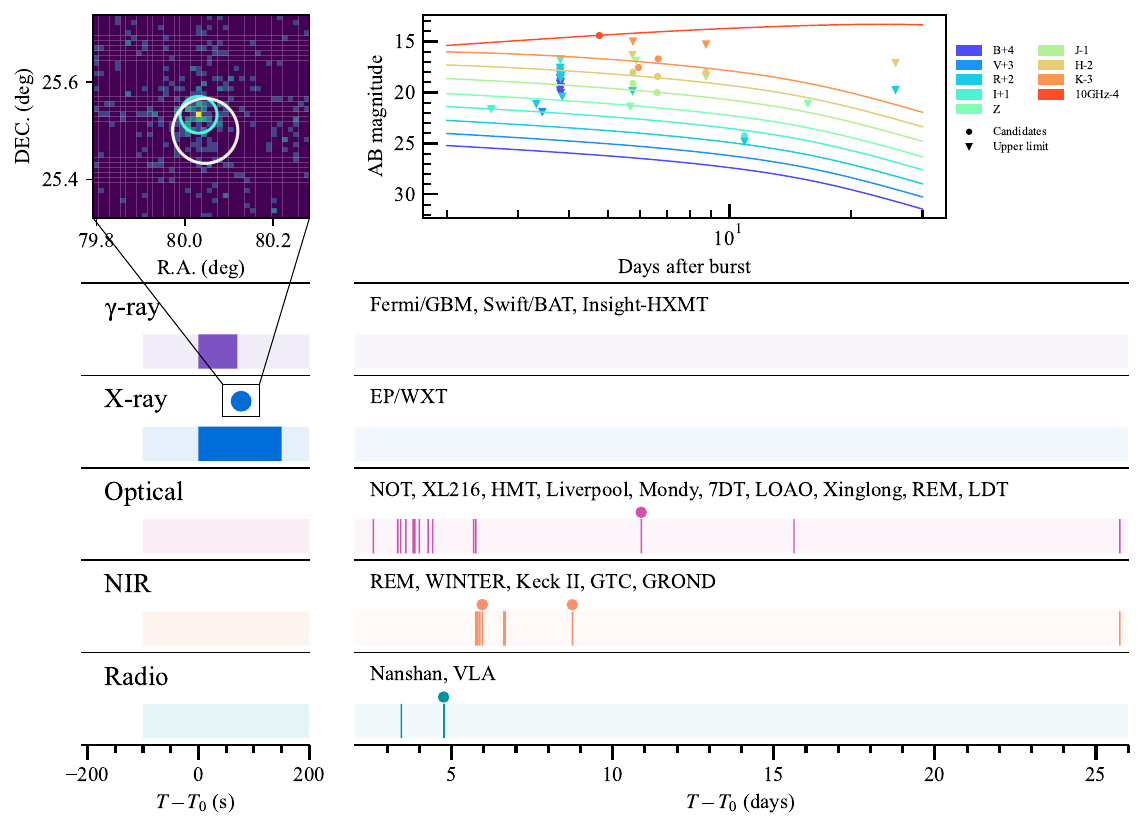}
 \caption{Multiwavelength observations of EP240219a/GRB 240219A. Each wavelength is represented by different colored dots, lines and blocks. The small dots indicate follow-up observations with candidates, while the lines denote the observing times; lines for follow-up observations are artificially thickened for clarity. The telescopes corresponding to each wavelength are listed in the same row. The top panels display EP source observation with the EP/WXT position marked with a light blue circle and the Swift/BAT position marked with a white circle and afterglow detections of Candidate 3 with upper limits. An unconstrained afterglow model (with a typical parameter set of $E_{\rm k,iso}=9.85\times10^{53}~\rm{erg}$, $\theta_{\rm v}=8.99\times10^{-2}\ \rm{rad}$, $\theta_{\rm c}=2.65\times10^{-2}\ \rm{rad}$, $n=2.69\times10^{-2}\ \rm{cm^{-3}}$, $p=2.94$, $\epsilon_{\rm e}=0.13$, $\epsilon_{\rm B}=4.11\times10^{-2}$, $v=0.28\ \rm{mag}$, $z=1.5$) is adopted for illustration purposes.}
 \label{fig:obs}
\end{figure*}

EP/WXT operates in the 0.5 - 4.0 keV soft X-ray band, offering an energy resolution of 130 eV (at 1.25 keV) and a temporal resolution of 0.05 s. The observation of GRB 240219A took place between 06:07:32.034 UT and 06:30:44.934 UT on 2024 February 19 resulting in a net exposure time of 1033 s. GRB 240219A was identified within the WXT's field of view at R.A. = 80.$^{\circ}$031, decl. = 25.$^{\circ}$533 with an uncertainty of 2.$'$3 arcminutes, as illustrated in Figure \ref{fig:obs}. The X-ray photon events were processed and calibrated using specialized data reduction software and the calibration database (CALDB) developed for EP/WXT (Y. Liu et al. in preparation). The generation of CALDB relies on the results from both on-ground and in-orbit calibration campaigns (H. Cheng et al. in preparation). Energy corrections were applied to each event employing bias and gain values stored in CALDB while identifying and flagging bad/flaring pixels. Events that met the predefined criteria, such as singlets, doublets, triplets, and quadruplets without anomalous flags, were selected to generate a cleaned event file. Subsequently, the image in the 0.5 – 4.0 keV energy range was extracted from the cleaned events, with each photon's position projected into celestial coordinates. The source and background light curves and spectra for any specified time interval were obtained from a circular region with a radius of 9.$'$1 and an annular region with inner and outer radii of 1.$^{\circ}$4 and 1.$^{\circ}$8, respectively. Additionally, as the instrument's in-orbit calibration was ongoing during the observation, the calibration uncertainties are accounted for in the joint spectral energy distribution (SED) fitting.

Following EP/WXT’s alert, we retrieved the time-tagged event dataset encompassing the temporal span of GRB 240219A from the publicly accessible Fermi/GBM data archive\footnote{\url{https://heasarc.gsfc.nasa.gov/FTP/fermi/data/gbm/daily/}}. Among the twelve sodium iodide detectors, n9 and na were chosen due to their minimal viewing angles relative to the GRB source direction. This direction is determined by the angle between the detector's pointing direction and the GRB location as identified by EP/WXT. The burst signal-to-noise ratio of the two selected detectors both exceed 3 $\sigma$. Furthermore, for both temporal and spectral analyses, we included the bismuth germanium oxide detector b1, which also has a relatively small viewing angle. Given the weak signal detected by b1, we evaluated the spectral fit quality both with and without its data and found that incorporating the data from b1 resulted in an improved fit. Subsequently, we analyzed the Fermi/GBM data using the established methodologies described in \cite{Zhang_2011} and \cite{2022Natur.612..232Y}.

\subsection{Light Curve}
\begin{figure}
 \centering
 \includegraphics[width = 0.45\textwidth]{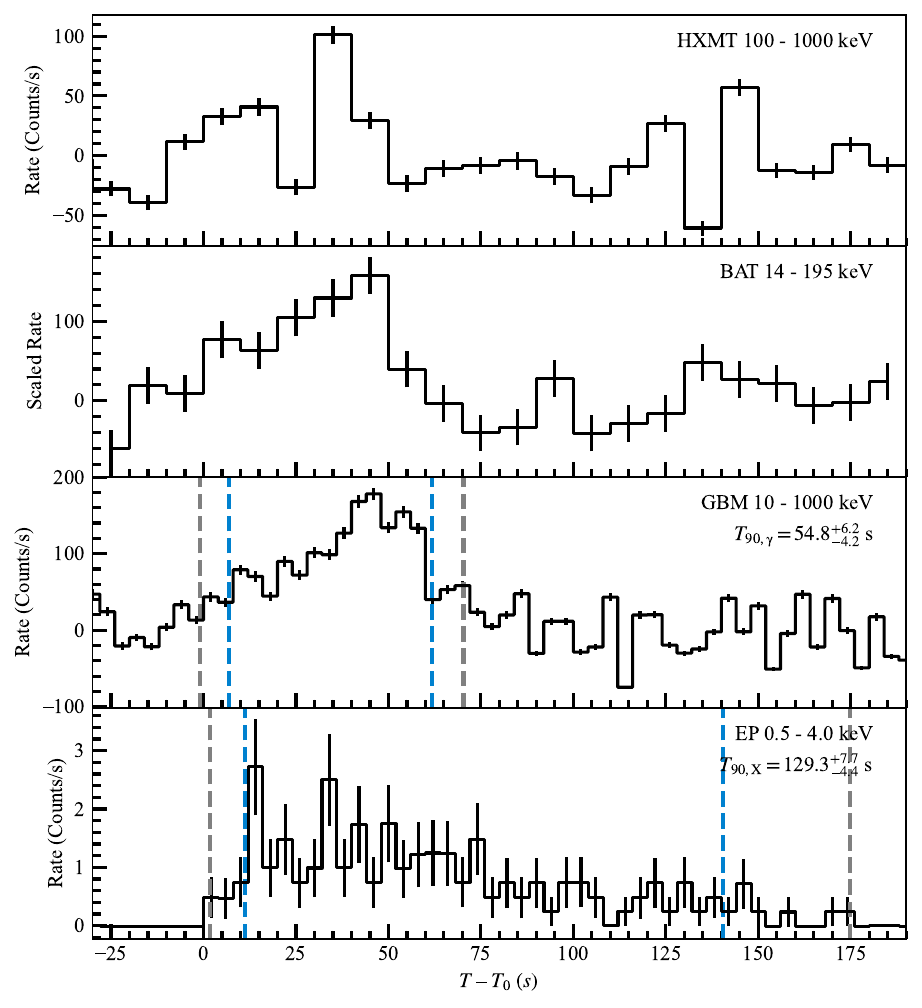}
 \caption{Light curves of GRB 240219A as observed by Insight-HXMT/HE, Swift/BAT, Fermi/GBM and EP/WXT. The top two panels show the light curve of Insight-HXMT/HE in the energy range of 100--1000 keV and Swift/BAT in the energy range of 14--195 keV with a bin size of 10 s. The third panel illustrates the light curve of Fermi/GBM in the energy range of 10--1000 keV, while the bottom panel displays the light curve of EP/WXT in the energy range of 0.5--4.0 keV. The bin size for Fermi/GBM and EP/WXT light curves is uniformly set to 4 s. The blue dashed vertical lines represent the $T_{\rm 90}$ interval, and the gray dashed vertical lines denote the $T_{\rm 100}$. All error bars mark the 1 $\sigma$ confidence level.}
 \label{fig:lc}
\end{figure}

In Figure \ref{fig:lc}, we present the light curves of GRB 240219A, featuring Fermi/GBM data with a bin size of 4 s in the energy range of 10--1000 keV and EP/WXT data in 0.5--4.0 keV. The light curves of Insight-HXMT/HE in the energy range of 100--1000 keV and Swift/BAT in the energy range of 14--195 keV are displayed with a bin size of 10 s. The Swift/BAT light curve is obtained and rebinned from the rates data in BAT science products. The start time and peak time of Insight-HXMT/HE and Swift/BAT detections are consistent with those of Fermi/GBM. Since the Fermi/GBM data provide a higher detection significance and broader energy range, we utilize only the Fermi/GBM data in the subsequent temporal and spectral analysis to ensure accuracy and simplicity. The overall profiles of X-ray and gamma-ray emissions exhibit coherence from the trigger time $T_{\rm 0}$, with the X-ray emission displaying a prolonged duration characterized by a slow decay. The duration of the burst, $T_{\rm 90, \gamma}$, in the energy range of 10--1000 keV is $54.8_{-4.2}^{+6.2}$ s, while in the energy range of 0.5--4.0 keV, the derived $T_{\rm 90, X}$ is $129.3_{-4.4}^{+7.7}$ s. 

The minimum variability timescale (MVT) is a measurement of the temporal variability of the light curves. We determined the MVT of GRB 240219A by implementing the Bayesian method \citep{Scargle_2013} to recognize Bayesian blocks from the time-tagged event data of Fermi/GBM and EP/WXT. We adopt the definition of MVT as half of the minimum bin size of the resulting blocks. For Fermi/GBM data, the MVT was found to be 2.19 s, whereas for EP/WXT data, it was 29.3 s. Consequently, the MVT of 2.19 s, derived from Fermi/GBM data, is considered as the MVT of this event. The MVT and gamma-ray luminosity in GRBs are found to exhibit an empirical anticorrelation \citep{2015ApJ...805...86S, 2023A&A...671A.112C}. Attributing the MVT to the viscous instability of the hyperaccretion disk, such a correlation can be well understood within the black hole central engine model \citep{2017ApJ...838..143X}. For GRB 240219A, the MVT and gamma-ray luminosity (assuming redshift $0.5 < z < 4$) are consistent with the empirical anticorrelation and theoretical predictions from the black hole central engine model.

The comparison of the Fermi/GBM and EP/WXT light curves shown in Figure \ref{fig:lc} highlights some temporal disparities in emission peak times and slightly varied structures between the two detections. This is further manifested by the lag calculation, which yields $-4.0^{+1.1}_{-1.1}$ s. The calculation of spectral lag was conducted using the method described in \citet{Zhang_2012}. This method employs the cross-correlation function \citep{2000ApJ...534..248N, Ukwatta_2010} to measure the time lags between the light curves from Fermi/GBM and EP/WXT shown in Figure \ref{fig:lc}. Considering that the bin size is 4 s, such a lag cannot be ruled out as being caused by noise fluctuations. Due to the weak detection of GRB 240219A by Fermi/GBM and the limited photons collected by EP/WXT, further discerning potential substructures of the emission is challenging and out of the scope of this study.

\subsection{Spectral Fit}
\label{subsec:specfit}
We conducted both time-integrated and time-resolved spectral fits for EP/WXT and Fermi/GBM data separately, covering the time range from 0 to 70 s. Time-integrated spectra for both instruments were extracted during this period. We divided the spectra of both instruments into several slices. However, to compare the spectral consistency, we required that the spectra of both instruments share the same corresponding time slices. To achieve this, we required a minimum of 20 total accumulating photon counts for each EP/WXT spectrum and 5 average accumulating photon counts per channel for each Fermi/GBM spectrum. We finally obtained the time-resolved spectra for three slices, as listed in Table \ref{tab:spec_fit}, for both instruments.

The fitting utilized the Python package MySpecFit, following the methodology summarized in \citet{2022Natur.612..232Y} and \citet{2023ApJ...947L..11Y}. MySpecFit is a Bayesian inference-based spectral fitting tool and uses the nested sampler Multinest \citep{Feroz2008MNRAS, Feroz2009MNRAS, Buchner2014A&A, Feroz2019OJAp} as the fitting engine. The evaluation of the goodness of fit involved examining the reduced statistic STAT/dof, where the term ``dof'' represents the degree of freedom and ``STAT'' serves as the likelihood function. Specifically, for the EP spectra, we used CSTAT \citep{Cash1979ApJ}, while for the GBM spectra, we used PGSTAT \citep{1996ASPC..101...17A}. The model comparison was conducted using the Bayesian information criterion (BIC) as outlined by \citet{bic_ref}. The fitting results of the best model and the corresponding statistics for each time slice are listed in Table \ref{tab:spec_fit}. From the spectral fittings, the spectral evolution and the SEDs were subsequently derived, as shown in Figure \ref{fig:spec_evo} and Figure \ref{fig:sed}. 

We briefly outline the key results of the spectral fitting as follows:

\begin{table*}
\centering
\small
\caption{Spectral fitting results and corresponding fitting statistics for EP/WXT and Fermi/GBM. All errors represent the 1 $\sigma$ uncertainties.}
\label{tab:spec_fit}
\begin{tabular}{cccccccccccccc}
\hline
\hline
\multicolumn{14}{c}{Independent Fit}\\
\hline
\multirow{3}{*}{t1 (s)} & \multirow{3}{*}{t2 (s)} & \multicolumn{12}{c}{PL Model}\\
\cline{3-14}
& & \multicolumn{3}{c}{$\alpha_{\rm X}$} & \multicolumn{3}{c}{${\rm log}A_{\rm pl}$} & \multicolumn{3}{c}{CSTAT/dof} & \multicolumn{3}{c}{BIC}\\
& & & & & \multicolumn{3}{c}{($\rm{photons\ cm^{-2}}\ s^{-1}\ keV^{-1}$)} & & & & & &\\
\hline
0.00 & 38.25 & \multicolumn{3}{c}{$-1.56_{-0.44}^{+0.38}$} & \multicolumn{3}{c}{$-2.90_{-0.80}^{+0.66}$} & \multicolumn{3}{c}{$10.95/10$} & \multicolumn{3}{c}{$15.92$}\\
38.25 & 53.21 & \multicolumn{3}{c}{$-1.76_{-0.65}^{+0.61}$} & \multicolumn{3}{c}{$-3.19_{-1.24}^{+1.08}$} & \multicolumn{3}{c}{$2.02/3$} & \multicolumn{3}{c}{$5.24$}\\
53.21 & 70.00 & \multicolumn{3}{c}{$-1.95_{-0.74}^{+0.70}$} & \multicolumn{3}{c}{$-3.64_{-1.39}^{+1.13}$} & \multicolumn{3}{c}{$5.73/3$} & \multicolumn{3}{c}{$8.95$}\\
70.00 & 160.00 & \multicolumn{3}{c}{$-2.63_{-0.63}^{+0.60}$} & \multicolumn{3}{c}{$-5.21_{-1.18}^{+1.05}$} & \multicolumn{3}{c}{$9.82/9$} & \multicolumn{3}{c}{$14.61$}\\
0.00 & 70.00 & \multicolumn{3}{c}{$-1.69_{-0.30}^{+0.25}$} & \multicolumn{3}{c}{$-3.11_{-0.53}^{+0.46}$} & \multicolumn{3}{c}{$18.34/21$} & \multicolumn{3}{c}{$24.61$}\\
\hline
\multirow{3}{*}{t1 (s)} & \multirow{3}{*}{t2 (s)} & \multicolumn{12}{c}{CPL Model}\\
\cline{3-14}
& & \multicolumn{2}{c}{$\alpha_{\rm \gamma}$} & \multicolumn{2}{c}{${\rm log}E_{\rm p,\gamma}$} & \multicolumn{4}{c}{${\rm log}A_{\rm cpl}$} & \multicolumn{2}{c}{PGSTAT/dof} & \multicolumn{2}{c}{BIC}\\
& & & & \multicolumn{2}{c}{(keV)} & \multicolumn{4}{c}{($\rm{photons\ cm^{-2}}\ s^{-1}\ keV^{-1}$)} & & & &\\
\hline
0.00 & 38.25 & \multicolumn{2}{c}{$-1.72_{-0.21}^{+0.35}$} & \multicolumn{2}{c}{$1.99_{-0.36}^{+0.63}$} & \multicolumn{4}{c}{$-3.21_{-0.18}^{+0.31}$} & \multicolumn{2}{c}{$308.42/352$} & \multicolumn{2}{c}{$326.03$}\\
38.25 & 53.21 & \multicolumn{2}{c}{$-0.93_{-0.37}^{+0.37}$} & \multicolumn{2}{c}{$2.21_{-0.12}^{+0.37}$} & \multicolumn{4}{c}{$-2.48_{-0.27}^{+0.23}$} & \multicolumn{2}{c}{$326.73/352$} & \multicolumn{2}{c}{$344.35$}\\
53.21 & 70.00 & \multicolumn{2}{c}{$-1.24_{-0.55}^{+0.08}$} & \multicolumn{2}{c}{$1.82_{-0.23}^{+0.63}$} & \multicolumn{4}{c}{$-2.84_{-0.51}^{+0.04}$} & \multicolumn{2}{c}{$317.37/352$} & \multicolumn{2}{c}{$334.99$}\\
0.00 & 70.00 & \multicolumn{2}{c}{$-1.40_{-0.28}^{+0.17}$} & \multicolumn{2}{c}{$2.11_{-0.14}^{+0.52}$} & \multicolumn{4}{c}{$-2.95_{-0.22}^{+0.13}$} & \multicolumn{2}{c}{$281.40/352$} & \multicolumn{2}{c}{$299.02$}\\
\hline
\hline
\\
\hline
\hline
\multicolumn{14}{c}{Joint Fit}\\
\hline
\multirow{3}{*}{t1 (s)} & \multirow{3}{*}{t2 (s)} & \multicolumn{12}{c}{CPL Model}\\
\cline{3-14}
& & \multicolumn{2}{c}{$\alpha$} & \multicolumn{2}{c}{${\rm log}E_{\rm p}$} & \multicolumn{2}{c}{${\rm log}A$} & \multicolumn{2}{c}{$f$} & \multicolumn{2}{c}{STAT/dof} & \multicolumn{2}{c}{BIC}\\
& & & & \multicolumn{2}{c}{(keV)} & \multicolumn{2}{c}{($\rm{photons\ cm^{-2}}\ s^{-1}\ keV^{-1}$)} & & & & & & \\
\hline
0.00 & 38.25 & \multicolumn{2}{c}{$-1.71_{-0.12}^{+0.06}$} & \multicolumn{2}{c}{$1.99_{-0.34}^{+0.43}$} & \multicolumn{2}{c}{$-3.21_{-0.14}^{+0.09}$} & \multicolumn{2}{c}{$1.05_{-0.12}^{+0.01}$} & \multicolumn{2}{c}{$(11.09+308.43)/363$} & \multicolumn{2}{c}{$343.14$} \\
38.25 & 53.21 & \multicolumn{2}{c}{$-1.45_{-0.07}^{+0.10}$} & \multicolumn{2}{c}{$2.61_{-0.28}^{+0.02}$} & \multicolumn{2}{c}{$-2.80_{-0.04}^{+0.09}$} & \multicolumn{2}{c}{$1.05_{-0.11}^{+0.02}$} & \multicolumn{2}{c}{$(3.51+330.07)/356$} & \multicolumn{2}{c}{$357.13$} \\
53.21 & 70.00 & \multicolumn{2}{c}{$-1.64_{-0.20}^{+0.07}$} & \multicolumn{2}{c}{$1.80_{-0.22}^{+0.59}$} & \multicolumn{2}{c}{$-3.15_{-0.27}^{+0.09}$} & \multicolumn{2}{c}{$1.04_{-0.09}^{+0.03}$} & \multicolumn{2}{c}{$(6.04+318.19)/356$} & \multicolumn{2}{c}{$347.78$} \\
0.00 & 70.00 & \multicolumn{2}{c}{$-1.70_{-0.05}^{+0.05}$} & \multicolumn{2}{c}{$2.41_{-0.32}^{+0.18}$} & \multicolumn{2}{c}{$-3.16_{-0.04}^{+0.08}$} & \multicolumn{2}{c}{$1.07_{-0.13}^{+0.01}$} & \multicolumn{2}{c}{$(18.43+282.98)/374$} & \multicolumn{2}{c}{$325.15$} \\
\hline
\hline
\end{tabular}
\end{table*}

\begin{enumerate}
\item \textit{The spectral models.} The best-fit models in our analyses are finally determined to be a PL for EP/WXT spectra and a cutoff power law (CPL) for Fermi/GBM spectra. CPL is also determined to be the best model for the joint EP/WXT and Fermi/GBM spectral fitting. We note that an absorption component is needed when EP/WXT data are involved.

\begin{figure}
 \centering
 \includegraphics[width = 0.45\textwidth]{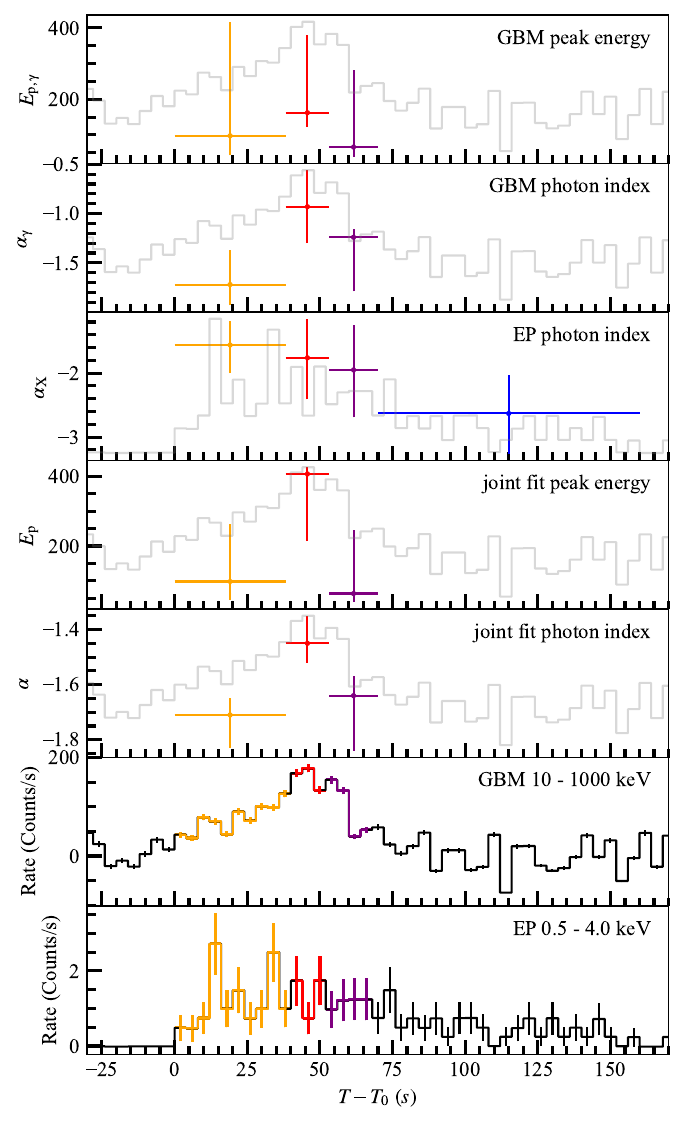}
 \caption{The observed light curves of GRB 240219A and its spectral evolution based on the best-fit parameters of the CPL model for Fermi/GBM spectra, the PL model for EP/WXT spectra, and the CPL model for Fermi/GBM-EP/WXT spectra. Same time slices are highlighted in same colors across all panels. All error bars on data points represent the 1 $\sigma$ confidence level.}
 \label{fig:spec_evo}
\end{figure}

\item \textit{The $N_{\rm{H}}$ value.} A fixed $N_{\rm{H}}$ value is adopted across all fits. To obtain this fixed $N_{\rm{H}}$ value, we first perform time-resolved spectral fits with linked $N_{\rm{H}}$ using an absorbed PL model \textit{tbabs*powerlaw} for the EP/WXT spectra, where \textit{tbabs} is the Tuebingen-Boulder interstellar medium absorption model \citep{Wilms2000ApJ} with a free parameter for the absorption column density, $N_{\rm{H}}$. The best-fit $N_{\rm{H}}$ is then adopted and fixed at $1.03_{-0.21}^{+0.39} \times 10^{22} \ \rm{cm^{-2}}$ in all subsequent spectral analyses.

\item \textit{Independent fit.}
\begin{itemize}
 \item[(a)] \textit{Time-integrated fit.} In the time interval of 0--70 s, the time-integrated spectrum of EP/WXT data can be well fitted by the absorbed PL model with an average photon index of $-1.69_{-0.30}^{+0.25}$. The spectrum of Fermi/GBM data is aptly characterized by the CPL model, featuring a typical GRB photon index of $-1.40_{-0.28}^{+0.17}$ and a peak energy of $127_{-32}^{+304}$ keV. Considering the larger error bars, the low-energy index of the CPL model is consistent with the photon index of the PL model, suggesting a common physical origin of the emissions in the two energy bands.
 
\begin{figure*}
 \centering
 \begin{tabular}{cc}
 \includegraphics[width=0.4\textwidth]{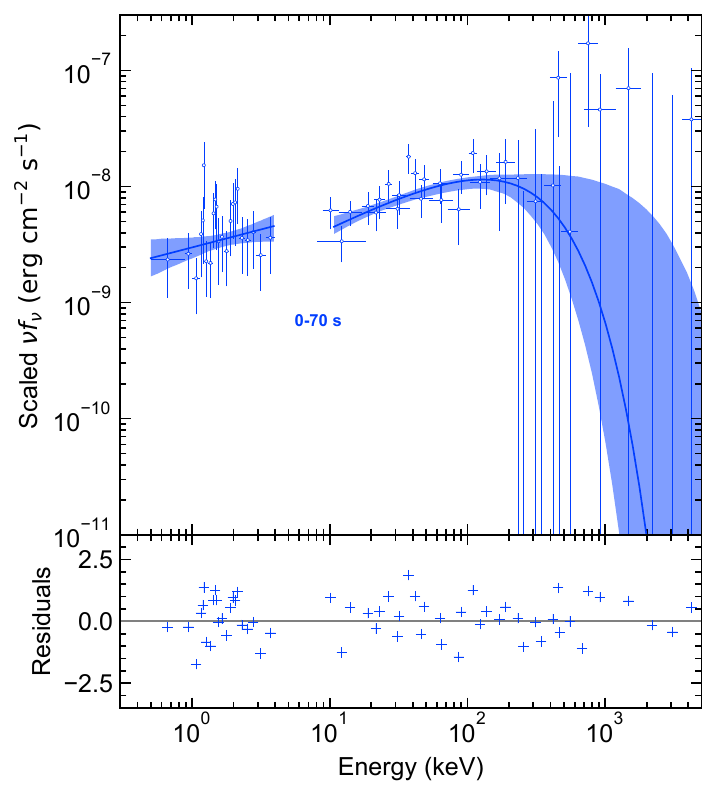} &
 \includegraphics[width=0.4\textwidth]{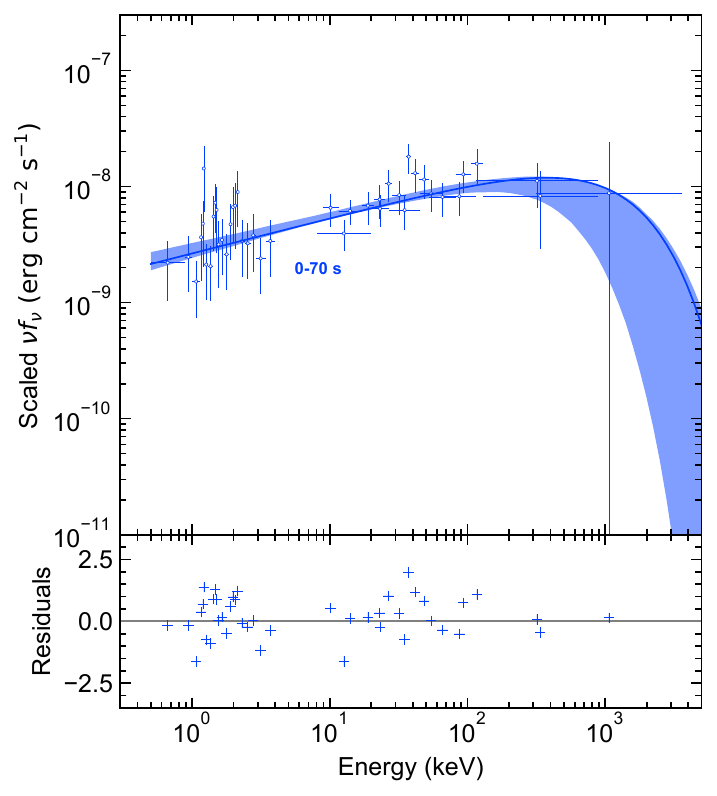} \\
 \includegraphics[width=0.4\textwidth]{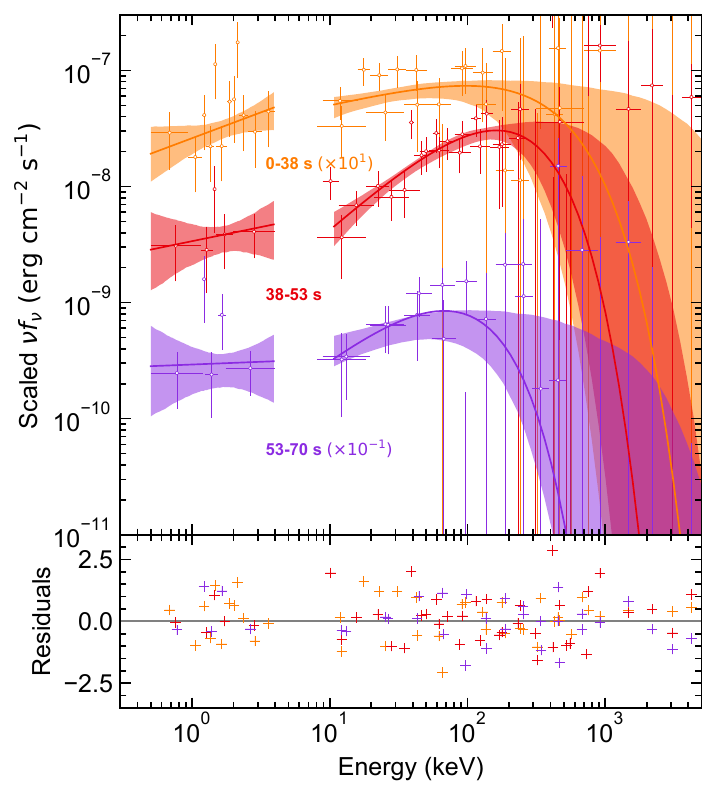} & 
 \includegraphics[width=0.4\textwidth]{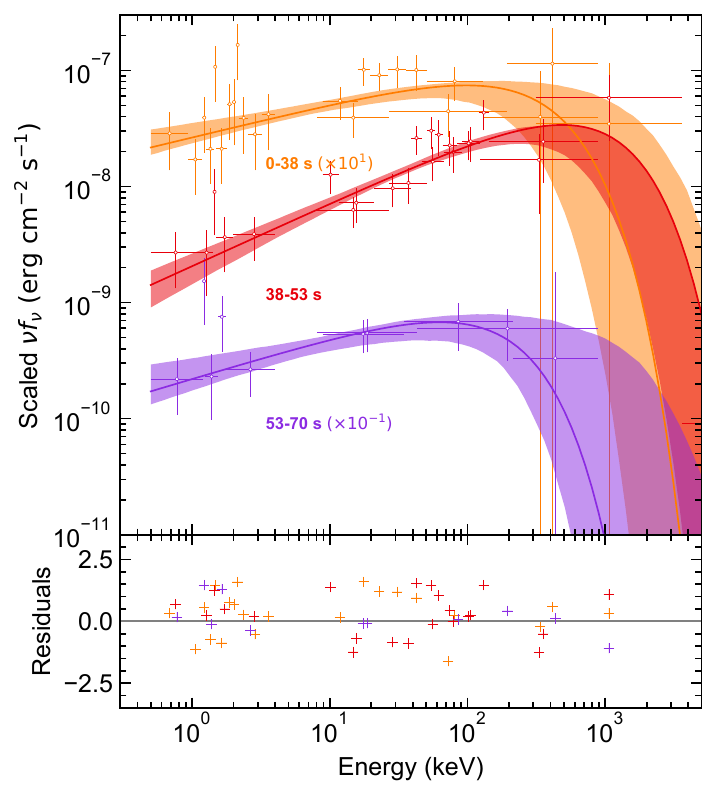} \\
   \end{tabular}
\caption{SEDs and their evolution. The left panels present SEDs derived from independent spectral fittings at different time intervals. Solid lines show the best-fit unabsorbed model for each independent fit. The right panels illustrate SEDs obtained from joint spectral fittings using an absorbed single CPL model at different time intervals. The unabsorbed CPL model is presented. Error bars represent the 1 $\sigma$ confidence level.}
 \label{fig:sed}
\end{figure*}

 \item[(b)] \textit{Time-resolved fit.} In the first time slice (0.00--38.25 s), both Fermi/GBM and EP/WXT spectra demonstrate a relatively lower peak energy and softer low-energy indices, indicating a soft spectrum emerging in the early phase of the burst. Afterwards, EP/WXT shows a trend of transition from hard to soft evolution. We note that the spectral indices in the second and third time slices of Fermi/GBM are not exactly aligned with EP/WXT (i.e., the photon indices are generally softer in X-ray spectra compared to gamma-ray spectra). However, in view of the large uncertainties caused by the limitation of photon numbers, the spectra of EP/WXT and Fermi/GBM are considered to be consistent within 1 $\sigma$ uncertainty.
\end{itemize}

\item \textit{Joint fit.} The joint fit is performed with an absorbed CPL model \textit{tbabs*cutoffpowerlaw}. Considering the calibration uncertainties for Fermi/GBM and EP/WXT, we introduced a calibration constant $f$ within the range of 90\%--110\% on EP/WXT to account for the systematic uncertainty between the two instruments. We found that for all four slices, the two components of the spectral data from EP/WXT and Fermi/GBM appear to be sewn together with a single CPL model, characterized by acceptable best-fit parameters (Table \ref{tab:spec_fit}), indicating a coherent single component spanning from X-ray to gamma-ray emissions. The joint fit also provides constraints on model parameters, which yields smaller uncertainties than what are obtained from independent fits, as is shown in Figure \ref{fig:sed}. Therefore, we adopt the best-fit model parameters of the joint fit for subsequent analyses. Additionally, the time-resolved joint fits indicate an initial soft spectrum followed by a hard-to-soft evolution, a trend that is not clearly manifested in the independent fits due to large uncertainties.
\end{enumerate}

\begin{table*}
\centering
\fontsize{7}{6.8}\selectfont
\caption{Multiwavelength Follow-up Observations of GRB 240219A. The first column lists the time elapsed after the EP trigger, adjusted by adding half of the exposure time of each observation. Correction for foreground Galactic extinction has not been applied to the magnitudes.}
\label{tab:obs}
\begin{tabular}{cccccccc}
\hline
\hline
$T-T_{\rm 0}$ (day) & $\Delta T (s)$ &(R.A., Decl.)& Telescope & Band & AB Magnitude & comment & Ref. \\
\hline
0.000 & - & (80.031, 25.533) & EP/WXT & x-ray & - && (1, 17) \\
0.000 & - & - & Fermi/GBM & gamma-ray & - && (2) \\
0.000 & 300 & (80.046, 25.500) & Swift/BAT & gamma-ray & - &&(3) \\
0.000 & - & - & Insight-HXMT/HE & gamma-ray & - && (4) \\
2.574 & 3$\times$120 & - & Liverpool/IO:O & $I$ & $\textgreater$ 22.12 && (5) \\
3.329 & 9$\times$200 & - & XL216 & $R$ & $\textgreater$ 21.3 && (6) \\
3.424 & 30$\times$90 & - & HMT & unfilter & $\textgreater$ 19.7 && (17) \\
3.443 & 12$\times$200 & - & Nanshan & $V$ & $\textgreater$ 21.7126 && (7) \\
3.582 & 9$\times$200 & (80.034, 25.548) & NOT & $z$ & 21.8$\pm$0.2 & candidate 0 (ruled out)& (17) \\
3.811 & 900 & - & 7DT & m400 & $\textgreater$ 18.166 && (8) \\
3.824 & 900 & - & 7DT & m425 & $\textgreater$ 18.338 && (8) \\
3.812 & 900 & - & 7DT & m450 & $\textgreater$ 18.330 && (8) \\
3.824 & 900 & - & 7DT & m475 & $\textgreater$ 18.447 && (8) \\
3.811 & 900 & - & 7DT & m500 & $\textgreater$ 18.597 && (8) \\
3.823 & 900 & - & 7DT & m525 & $\textgreater$ 18.678 && (8) \\
3.816 & 900 & - & 7DT & m550 & $\textgreater$ 18.379 && (8) \\
3.828 & 900 & - & 7DT & m575 & $\textgreater$ 18.486 && (8) \\
3.813 & 900 & - & 7DT & m600 & $\textgreater$ 17.712 && (8) \\
3.825 & 900 & - & 7DT & m625 & $\textgreater$ 18.365 && (8) \\
3.812 & 900 & - & 7DT & m650 & $\textgreater$ 18.110 && (8) \\
3.824 & 900 & - & 7DT & m675 & $\textgreater$ 18.425 && (8) \\
3.812 & 900 & - & 7DT & m700 & $\textgreater$ 17.894 && (8) \\
3.824 & 900 & - & 7DT & m725 & $\textgreater$ 18.029 && (8) \\
3.812 & 900 & - & 7DT & m750 & $\textgreater$ 17.154 && (8) \\
3.825 & 900 & - & 7DT & m775 & $\textgreater$ 17.641 && (8) \\
3.815 & 900 & - & 7DT & m800 & $\textgreater$ 17.177 && (8) \\
3.827 & 900 & - & 7DT & m825 & $\textgreater$ 17.014 && (8) \\
3.861 & 60$\times$30 & - & LOAO & $R$ & $\textgreater$ 20.562 && (9) \\
4.278 & 30$\times$120 & - & Mondy & $R$ & $\textgreater$ 18.00 && (10) \\
4.278 & 30$\times$120 & - & Mondy & $R$ & $\textgreater$ 18.63 && (10) \\
4.416 & 30$\times$90 & - & HMT & unfilter & $\textgreater$ 19.5 && (17) \\
4.765 & 2070 & (80.046, 25.560) & VLA & $X$ & 169 $\rm \mu Jy$ & candidate 1 (unlikely) & (11) \\
4.765 & 2070 & (80.050, 25.535) & VLA & $X$ & 82 $\rm \mu Jy$ & candidate 2 (unlikely) & (11) \\
4.765 & 2070 & (80.020, 25.530) & VLA & $X$ & 157 $\rm \mu Jy$ & candidate 3 (undetermined) & (11) \\
4.765 & 2070 & (80.025, 25.522) & VLA & $X$ & 38 $\rm \mu Jy$ & candidate 4 (undetermined)& (11) \\
4.765 & 2070 & (79.999, 25.546) & VLA & $X$ & 49 $\rm \mu Jy$ & candidate 5 (undetermined) & (11) \\
4.765 & 2070 & (80.020, 25.558) & VLA & $X$ & 30 $\rm \mu Jy$ & candidate 6 (ruled out) & (11) \\
4.765 & 2070 & (79.969, 25.530) & VLA & $X$ & 300 $\rm \mu Jy$ & candidate 7 (ruled out) & (11) \\
5.685 & 12$\times$200 & - & NOT & $z$ & $\textgreater$ 22.5 && (17) \\
5.754 & 1200 & - & REM & $r$ & $\textgreater$ 19.9 && (12) \\
5.754 & 1200 & - & REM & $H$ & $\textgreater$ 18.69 && (12) \\
5.768 & 1200 & - & GROND & $J$ & $20.80 \pm 0.37$ & candidate 3 & (17) \\
5.768 & 1200 & - & GROND & $H$ & $20.44 \pm 0.34$ & candidate 3 & (17) \\
5.768 & 1200 & - & GROND & $K_s$ &$ \textgreater 18.24$ & candidate 3 & (17) \\
5.877 & - & - & WINTER & $J$ & $\textgreater$ 18.5 & unknown exposure time & (13) \\
5.959 & 4$\times$300 & (80.020, 25.530) & Keck II/NIRES & $K^\prime$ & 20.84 & candidate 3 & (14) \\
6.619 & 91$\times$10 & (80.020, 25.530) & GTC/EMIR & $J$ & 21.66 $\pm$ 0.10 & candidate 3 & (17) \\
6.619 & 91$\times$10 & (80.025, 25.522) & GTC/EMIR & $J$ & $\textgreater$ 25.86 & candidate 4 & (17) \\
6.619 & 91$\times$10 & (79.999, 25.546) & GTC/EMIR & $J$ & 22.33 $\pm$ 0.11 & candidate 5 & (17) \\
6.639 & 188$\times$6 & (80.020, 25.530) & GTC/EMIR & $H$ & 20.86 $\pm$ 0.10 & candidate 3 & (17) \\
6.639 & 188$\times$6 & (80.025, 25.522) & GTC/EMIR & $H$ & $\textgreater$ 25.92 & candidate 4 & (17) \\
6.639 & 188$\times$6 & (79.999, 25.546) & GTC/EMIR & $H$ & 21.58 $\pm$ 0.09 & candidate 5 & (17) \\
6.665 & 420$\times$3 & (80.020, 25.530) & GTC/EMIR & $K_s$ & 19.99 $\pm$ 0.13 & candidate 3 & (17) \\
6.665 & 420$\times$3 & (80.025, 25.522) & GTC/EMIR & $K_s$ & $\textgreater$ 25.92 & candidate 4 & (17) \\
6.665 & 420$\times$3 & (79.999, 25.546) & GTC/EMIR & $K_s$ & 20.92 $\pm$ 0.13 & candidate 5 & (17) \\
8.748 & 1200 & - & GROND & $J$ & $\textgreater 20.06$ & candidate 3 & (17) \\
8.748 & 1200 & - & GROND & $H$ & $20.51 \pm 0.34$ & candidate 3 & (17) \\
8.748 & 1200 & - & GROND & $K_s$ & $\textgreater 18.53$ & candidate 3 & (17) \\
10.893 & 10$\times$150 & (80.020, 25.530) & LDT/LMI & $r^\prime$ & $\textgreater$ 24.9 & candidate 3 & (15)\\ 
10.893 & 10$\times$150 & (80.020, 25.530) & LDT/LMI & $i^\prime$ & 24.8 $\pm$ 0.2 & candidate 3 & (15)\\ 
10.893 & 10$\times$150 & (80.025, 25.522) & LDT/LMI & $r^\prime$ & $\textgreater$ 25.1 & candidate 4 & (15)\\ 
10.893 & 10$\times$150 & (80.025, 25.522) & LDT/LMI & $i^\prime$ & $\textgreater$ 24.5 & candidate 4 & (15)\\ 
10.893 & 10$\times$150 & (79.999, 25.546) & LDT/LMI & $r^\prime$ & $\textgreater$ 25.1 & candidate 5 & (15)\\ 
10.893 & 10$\times$150 & (79.999, 25.546) & LDT/LMI & $i^\prime$ & 24.7 $\pm$ 0.2 & candidate 5 & (15)\\ 
15.627 & 16$\times$200 & - & NOT & $z$ & $\textgreater$ 22.2 && (17) \\
25.741 & 1800 & - & REM & $r$ & $\textgreater$ 19.8 && (16) \\ 
25.741 & 1800 & - & REM & $H$ & $\textgreater$ 19.49 && (16) \\ 
\hline
\end{tabular}
\tabletypesize{\scriptsize}
\tablecomments{
(1) \citet{2024GCN.35773....1Z}; (2) \citet{2024GCN.35776....1F}; (3) \citet{2024GCN.35784....1D}; (4) \citet{2024GCN.35785....1C}; (5) \citet{2024GCN.35783....1B}; (6) \citet{2024GCN.35795....1X}; (7) EP mail list; (8) \citet{2024GCN.35790....1P}; (9) \citet{2024GCN.35791....1P};  (10) \citet{2024GCN.35787....1P}; (11) \citet{2024GCN.35788....1H}; (12) \citet{2024GCN.35803....1F}; (13) \citet{2024GCN.35807....1K}; (14) \citet{2024GCN.35808....1K}; (15) \citet{2024GCN.35865....1S}; (16) \citet{2024GCN.35978....1F}; (17) this work.
}
\end{table*}

\section{Long-term follow-up and afterglow candidates}
\label{sec:obs}
Follow-up observations were carried out spanning various wavelengths, from optical to radio. Since the onboard trigger was not activated during the EP commissioning phase, the first follow-up observations started approximately 2.6 days after the burst trigger. A summary of these observations, detailing the candidates and their respective upper limits, is shown in Table \ref{tab:obs} and Figure \ref{fig:obs}.

Rapid responses and extensive efforts to search for the afterglow of GRB 240219A were made by the Liverpool Telescope, the Beijing Faint Object Spectrograph and Camera mounted on the Xinglong 2.16 m Telescope \citep[][XL216;]{Fan2016PASP}, the Half-meter Telescope (HMT; located at Xingming Observatory, China), the Nanshan Telescope, the 7-Dimensional Telescope (7DT) and the Lemonsan Optical Astronomy Observatory (LOAO). However, only upper limits were obtained during the first 3.5 days. Additionally, adverse weather conditions and poor seeing affected observations with the Gran Telescopio CANARIAS (GTC) and Lijiang Telescope. Unfortunately, despite these comprehensive early efforts, no optical or radio afterglow candidates were discovered.

At $\sim T_{\rm 0}$ + 3.6 days, the first reported afterglow candidate was detected by the efforts of the EP collaboration using the Nordic Optical Telescope (NOT) in the $z$-band. On the first night of NOT observations, we found an uncatalogued target with a brightness of $z=21.8\pm0.2$ at equatorial coordinates (J2000.0): R.A. = $05^h 20^m 08.^s24$ and decl. = $+25^{\circ} 32^{'} 50.83^{''}$. The magnitude of the candidate seemed to increase slightly during the second night of observations. To make a final check of this burst, we applied for one more observation on NOT at $\sim\ T_{\rm 0}$ + 15.6 days. The candidate we found before was still bright, which is not consistent with the GRB afterglow model. We also stacked all the NOT images and did not find any additional candidate associated with EP240219a. The photometric results of these observations were calibrated with the Pan-STARRS DR2 catalog \citep{Chambers2016arXiv,Flewelling2020ApJS} and are presented in Table \ref{tab:obs}.

On 2024 February 24, at 00:26 UTC, the Very Large Array (VLA) started observing the location of GRB 240219A \citep{2024GCN.35788....1H}. During 34.5 minutes of on-source time in the $X$-band (10 GHz) radio frequency, seven point sources were detected as potential afterglow candidates within the 3$'$ radius region. Among these, Candidates 6 and 7 were cross-matched with ZTF deep reference $r$-band images. However, the identification of Candidate 7 in an earlier VLA Sky Survey indicates that it is unlikely to be linked with GRB 240219A. Later on, Candidate 6 was ruled out due to a lack of variations from the observation of NIRES acquisition camera on the Keck II telescope. Candidate 1 and 2 were ruled out because of the optical extended source with low i$^\prime$ magnitude from the observation of the Large Monolithic Imager (LMI) on the 4 m Lowell Discovery Telescope (LDT).

We conducted our near-infrared follow-up using the Espectrógrafo Multiobjeto Infra-Rojo (EMIR) at the GTC starting from 21:04:48.89 UTC on February 25. Observations utilized ${J}$-, ${H}$-, and ${K_s}$-band filters. Standard data reduction, including image combination, distortion correction, and flat fielding, was performed using the Pyemir software package \cite{Pascual+10}. For astrometric solutions, 60 bright, unsaturated stars from the 2MASS point source catalog were selected as Web Coverage Service reference stars \citep{2mass+2003}. Coordinates were measured using the \textit{imexamine} task of the Image Reduction and Analysis Facility \citep[IRAF,][]{IRAF1,IRAF2}.\footnote{IRAF is distributed by the National Optical Astronomy Observatory, operated by the Association of Universities for Research in Astronomy under a cooperative agreement with the National Science Foundation.} The \textit{CCMAP} routine in IRAF was used for precise astrometric solutions, with rms uncertainties in (R.A., decl.) of ($0.''085$, $0.''059$) for all frames. Six candidates were identified in the GTC field. Without observing standard stars, we selected 20 bright, isolated, nonsaturated stars from the 2MASS catalog within the GTC field as photometric reference stars for each band. The zero point of the images was established with these reference stars, and apparent magnitudes were obtained using the background-subtracted flux of the candidates and the zero-point. The final AB magnitudes, converted using the Vega and AB magnitude relationship from \cite{Blanton+2005AJ}, are listed in Table \ref{tab:obs}.

Further additional detection of Candidate 3 was made using the Gamma-ray Burst Optical Near-infrared Detector \citep[GROND;][]{2008PASP..120..405G} at the 2.2m MPG telescope at La Silla at $T_0 + 8.748$ days. This detection showed a slightly higher flux density compared to the GTC observation at $T_0 + 6.639$ days. Additionally, the $K_s$ band detection of Candidate 3 from GTC at $T_0 + 6.665$ days is brighter than the detection from Keck II at $T_0 + 5.959$ days. However, due to the limited number of observations for Candidate 3, we are currently unable to confirm whether it is the afterglow counterpart or to discuss additional components/features, such as a supernova.

As of this writing, all follow-up observations did not confirm any candidate as the afterglow counterpart associated with GRB 240219A. Candidates 3, 4 and 5 remain undetermined. The absence of confirmed afterglow counterpart and the current uncertainty are primarily influenced by several factors: the relatively long latency ($\sim$ 44.5 hr) of the burst notice \citep[EP ATel;][]{2024ATel16463....1Z, 2024ATel16472....1Z}, which caused a relatively late start of follow-up observations; the large localization region initially provided by EP/WXT during its commissioning phase, which made the search for optical sources more challenging; and the possibility of an intrinsically dark afterglow and/or a high redshift of the burst. In the latter case, according to the Amati relation (see Section \ref{sec:nature}), the redshift of the burst may exceed $\sim$ 1.5, thus explaining the absence of detectable afterglow at $T_{\rm 0} + 2.6$ days and beyond.

\section{Nature of the Burst}
\label{sec:nature}
\begin{figure}
 \centering
 \includegraphics[width = 0.45\textwidth]{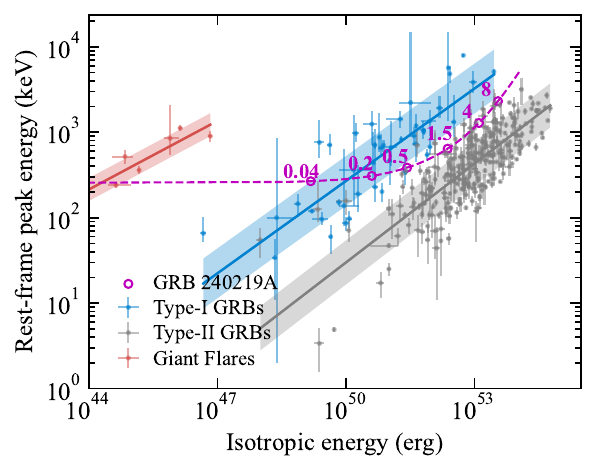}
 \caption{The $E_{\rm p,z}$-$E_{\rm iso}$ diagram. The blue, gray, and red solid lines represent the best-fit correlations for Type-I, Type-II, and magnetar giant flare populations, respectively. The purple dashed line indicates the position of GRB 240219A at various redshifts, with specific redshifts marked by dots. Error bars on data points represent the 1 $\sigma$ confidence level.}
 \label{fig:epeiso}
\end{figure}

As a bright X-ray transient detected by EP/WXT, EP240219a's reclassification as a GRB is based on the simultaneous gamma-ray detection, characterized by a hard spectrum and peak energy exceeding 100 keV. Due to the unknown distance of this cosmological event, we plot this GRB along the curve with various redshifts on the $E_{\rm p,z}$-$E_{\rm iso}$ diagram \citep{2002A&A...390...81A}. Based on Figure \ref{fig:epeiso}, we deduce that for GRB 240219A to appear as an intrinsic Type-II GRB, its redshift is likely no less than approximately 1.5. Furthermore, the detection by EP/WXT may indicate significant X-ray emission, providing an opportunity to investigate the subclassification of the GRB and the dominated spectral components during the prompt emission phase.

\subsection{Classical Gamma-ray Busrt, XRR or X-ray flash?}
Within the broad sample of GRBs, which exhibit a wide range of peak energies, two empirically identified subclasses with more dominant X-ray emission are known as X-ray flashes (XRFs) and XRRs \citep{10.1007/10853853_4, 10.1007/10853853_5, 2003A&A...400.1021B, 2004ASPC..312...12A, 2005A&A...440..809B, 2007A&A...461..485S}. XRFs are characterized by stronger X-ray emission compared to the typical intensity observed in classical gamma-ray bursts (C-GRBs). XRRs occupy an intermediate position between XRFs and C-GRBs, displaying relatively softer gamma-ray emission than C-GRBs. The classification of these subclasses has been systematically studied in various papers \citep{2003AIPC..662..244K, Sakamoto_2005, 2006A&A...460..653D, Sakamoto_2006, Sakamoto_2008, Bi_2018, Liu_2019}. However, the exact physical processes underpinning the manifestation of XRFs and XRRs remain contentious within the field.
\begin{figure}
 \centering
 \includegraphics[width = 0.45\textwidth]{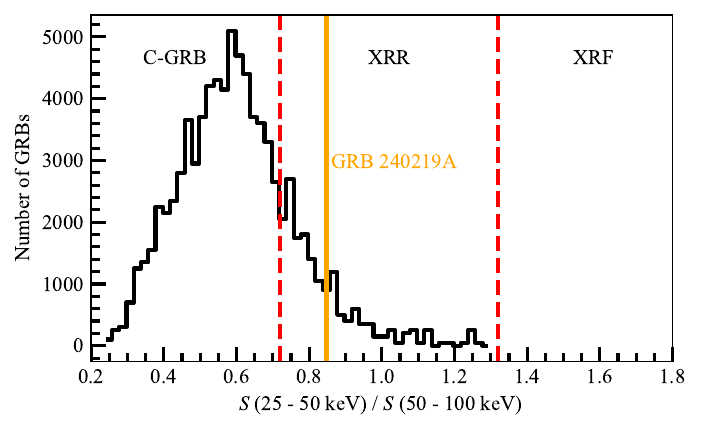}
 \includegraphics[width = 0.45\textwidth]{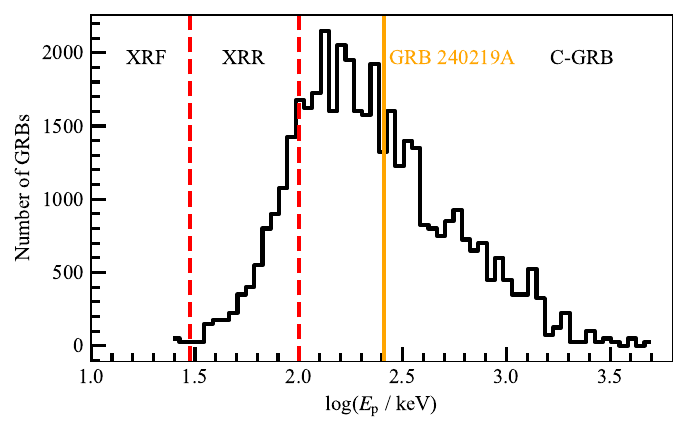}
\caption{Fluence ratio and $E_{\rm p}$ distribution of Fermi/GBM-detected GRB samples. The red dashed vertical lines denote boundaries of GRB subclasses, labeled accordingly. Yellow vertical lines indicate the location of GRB 240219A on the plot.}
 \label{fig:sample}
\end{figure}
\begin{figure}
 \centering
 \includegraphics[width = 0.45\textwidth]{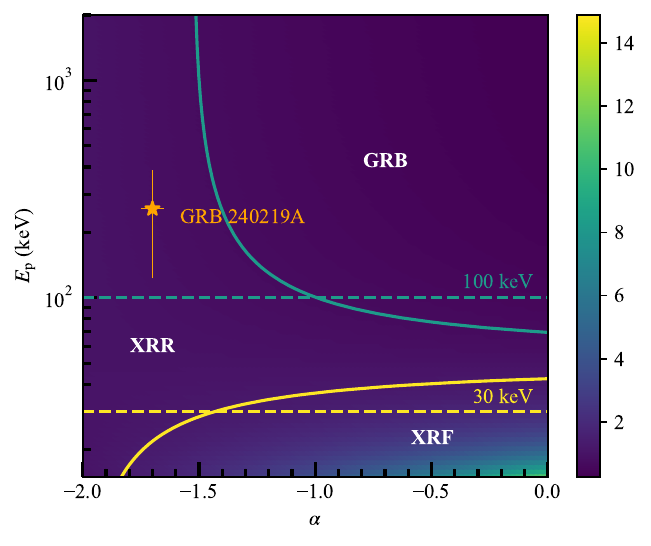}
\caption{Fluence ratio distribution in the parameter space of the CPL model. The $x$- and $y$-axes represent the photon index and peak energy from the CPL model, respectively, with a color gradient indicating the corresponding fluence ratio. Solid curves delineate boundaries for fluence ratio subclassification criteria ($S(25 - 50\ \rm{keV})$ / $S(50 - 100\ \rm{keV})$), while dashed horizontal lines indicate peak energy criteria. The location of GRB 240219A is marked with an orange star, with the error bar representing 1 $\sigma$ confidence level .}
 \label{fig:subclass}
\end{figure}

To classify C-GRBs, XRRs, and XRFs, two approaches based on spectral properties can be adopted. Initially, XRFs are identified through detection by the BeppoSAX/Wide Field Camera but remain untriggered by the BeppoSAX/Gamma Ray Burst Monitor \citep{10.1007/10853853_4}. The spectral peak energy, $E_{\rm p}$, is then used as a criterion to distinguish between these subclasses \citep{Sakamoto_2005, 2018pgrb.book.....Z}. \citet{Sakamoto_2005} established the boundary $E_{\rm p}$ at 30 keV between XRFs and XRRs and at 100 keV between XRRs and C-GRBs \citep{2018pgrb.book.....Z}. Alternatively, considering cross-instrument compatibility, C-GRBs, XRRs, and XRFs are proposed to be categorized according to the fluence ratio in different bands, such as $S(25 - 50\ \rm{keV})$ / $S(50 - 100\ \rm{keV})$ \citep{Sakamoto_2008} and $S(2 - 30\ \rm{keV})$ / $S(30 - 400\ \rm{keV})$ \citep{Sakamoto_2005}.

To explore the subclassification of GRB 240219A, we utilize the best-fit parameters of the time-integrated joint-fit spectrum with one CPL model obtained in Section \ref{subsec:specfit}. The observed peak energy of the burst aligns with the typical values of Fermi/GBM detected GRB samples\footnote{\url{https://heasarc.gsfc.nasa.gov/W3Browse/fermi/fermigbrst.html}}, categorizing it as a C-GRB. However, the fluence ratio, $S(25 - 50\ \rm{keV})$ / $S(50 - 100\ \rm{keV})$, is $0.85_{-0.10}^{+0.15}$, which deviates from the expected values for C-GRBs. This indicates that the burst is classified as an XRR, as illustrated in Figure \ref{fig:sample}. We note that the other fluence ratio, introduced by \citet{Sakamoto_2005} in the study of HETE-2 GRB samples, log[$S(2 - 30\ \rm{keV})$ / $S(30 - 400\ \rm{keV})$] $=-0.25 \pm 0.06$, also classifies it as an XRR.

We further investigated the boundaries of the fluence ratio $S(25 - 50\ \rm{keV})$ / $S(50 - 100\ \rm{keV})$ within the parameter space of the CPL model. As depicted in Figure \ref{fig:subclass}, there is a strong correlation between the fluence ratio and the observed spectral peak energy, as previously observed by \citet{Sakamoto_2005}, particularly when the photon index $\alpha$ exceeds $\sim$ -1.2. However, GRB 240219A exhibits a lower value of $\alpha$, placing it in a controversial region where the fluence ratio fails to constrain a reasonable peak energy. This event challenges the current working definitions of GRB subclassification, highlighting the need for further exploration into the subclass nature of GRBs using a larger sample of events detected across a broad energy range, such as the soft X-rays of prompt emission detected by EP/WXT.

\subsection{Implication of spectral components}
The GRB prompt emission spectrum often comprises three elemental spectral components, as discussed by \citet{Zhang_2011}: a nonthermal component, a quasi-thermal component, and another nonthermal high-energy component. While the third component remains undetected in GRB 240219A, the first component, well described by the CPL model, dominates throughout the burst. Here, as the photon indices $\alpha$ fall below the so-called synchrotron death line \citep[-2/3;][]{1998ApJ...506L..23P}, we attribute the CPL component to the nonthermal synchrotron radiation in the optically thin region. Moreover, to establish a lower limit for the anticipated photosphere spectrum within the framework of the internal shock model in the baryon-dominated outflow, we first assume that the photosphere emission of the ejecta generates a blackbody spectrum. By comparing how such a blackbody spectrum fits the observed data, we can measure the extent to which the outflow deviates from a purely thermal photospheric origin.

\begin{figure}
 \centering
 \includegraphics[width = 0.45\textwidth]{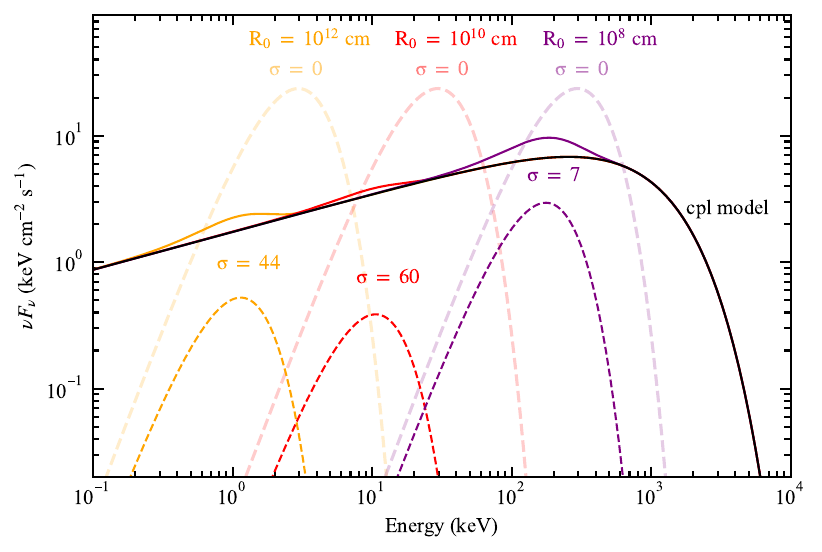}
 \caption{Calculation of the Poynting luminosity and baryonic luminosity ratio $\sigma$. The time-integrated spectrum of GRB 240219A is represented by the black solid curve. Colored dashed curves indicate synthetic blackbody components. Semitransparent thick curves denote components when $\sigma$ values are zero, while thin curves represent components with the lower limit value of $\sigma$, beyond which photosphere emission is not detectable. Colored solid curves show the hybrid model combining the CPL component (black solid line) and blackbody components (colored dashed lines) with the lower limit value of $\sigma$. The fireball base radius, $R_0$, is assumed to be three different values: $10^8$, $10^{10}$ and $10^{12}$ cm, marked in purple, red, and yellow, respectively. The derived lower limit values of $\sigma$ for these radii are 7, 60, and 44, respectively.}
 \label{fig:components}
\end{figure}

We examine the significance of the presence of a quasi-thermal-like component in the observed spectrum of GRB 240219A. As studied in detail by \citet{Gao_2015}, the theory of photospheric emission from a hybrid relativistic outflow allows us to derive the observed temperature $T_{\rm obs}$ and thermal flux $F_{\rm BB}$ of the photospheric emission from a given set of central engine parameters $(L_w, R_0, \eta, \sigma)$. In this framework, assuming a radiative efficiency of 50\%, the wind luminosity is given by $L_w = 2 L_{\rm obs}$ \citep{Gao_2015, Song_2024, 2024ApJ...972....1L, icmart}. We consider different values for the jet base radius $R_0$, specifically $10^8$, $10^{10}$, and $10^{12}$ cm. The dimensionless entropy $\eta$ and magnetization parameter $\sigma \equiv L_p/L_b$, where $L_p$ is the Poynting luminosity and $L_b$ is the baryonic luminosity, are two critical parameters governing the central engine dynamics. When $\eta \gg 1$ and $\sigma \ll 1$, a dominant thermal photospheric emission component is expected. Therefore, we set $\eta = 10^5$ and $\sigma = 10^{-5}$ to generate the pure hot fireball component spectra. Figure \ref{fig:components} shows that significant thermal-like peaks deviate from the observed spectrum.

Our analysis suggests a significant magnetization of the central engine of GRB 240219A, with the majority of energy being carried by magnetic fields rather than photons in the hot outflow. This phenomenon effectively suppresses the bright photospheric component. Furthermore, we can estimate a minimum value of the magnetization parameter $\sigma$, which represents the fraction of baryonic flux concealed. To estimate the lower limit of $\sigma$, we implement the following approach \citep[][see also \citealt{Gao_2015}]{2023ApJ...947L..11Y}. First, we construct a hybrid model by artificially adding a blackbody component, representing nondissipative photosphere emission, to the CPL model. In this hybrid model, the parameters $T_{\rm obs}$ and $F_{\rm BB}$ are expressed as functions of $\eta$ and $\sigma$ \citep{Gao_2015}. Additionally, the CPL model parameters are fixed at the best-fit values obtained from fitting GRB 240219A using the CPL-only model. In the study of hybrid jet components from a Fermi/GBM sample, the dimensionless entropy $\eta$ for all bursts exhibits an average value of $\sim 10^3$ \citep{Li_2020}. Therefore, we adopt $\eta = 10^3$ in our hybrid model to constrain the value of $\sigma$. We then determine the minimum $\sigma$ value at which the goodness of fit to the data becomes unacceptable. To quantify this, we employ the Akaike information criterion \citep[AIC;][]{1100705, doi:10.1080/03610927808827599}, setting the threshold for the goodness of fit at $\Delta \mathrm{AIC} > 5$ \citep{Krishak_2020}. Our calculations reveal a lower limit of $\sigma \geq 7$ for $R_0 = 10^8~\mathrm{cm}$, indicating that the outflow is mostly dominated by Poynting flux.

\section{Summary and Discussion}
\label{sec:sum}
In this Letter, we report on the first EP detection of a bright X-ray flare, EP240219a, which is associated with an untriggered GRB with consistent trigger time and overall profile. The peak of the burst shows a delay between the detections by EP/WXT and Fermi/GBM. Nonetheless, the possible substructures are not resolved because of the weak gamma-ray detection and the limited number of X-ray photons. We then conducted both individual and joint spectral fits with EP/WXT and Fermi/GBM data. Our findings suggest a seeming spectral discrepancy between the two emission bands when fitted individually. However, due to the large uncertainties, this discrepancy is likely not significant. A single CPL model can sufficiently account for the overall spectra.
This indicates a coherent broad emission of a typical GRB with a photon index of $-1.70_{-0.05}^{+0.05}$ and a peak energy of $257_{-134}^{+132}$ keV. Attributing the CPL component to the nonthermal synchrotron radiation, we further obtained a lower limit of the Poynting flux and the baryonic flux ratio $\sigma \geq 6$, implying the outflow is predominantly Poynting flux dominated. Furthermore, the spectral characteristics contribute additional insights into the subclassification of GRBs. Identifying GRB 240219A as an XRR with a high peak energy presents both challenges and opportunities for understanding and exploring the physical conditions of XRFs, XRRs, and C-GRBs.

Long-term follow-up observations are essential to build on the findings from GRB 240219A. Although localization and follow-up detection efforts did not yield a confirmed afterglow, they highlight the need for rapid response to capture the precise location and early afterglow phases, which are critical for progenitor and central engine studies.

The observation of GRB 240219A showcases EP's ability to detect and analyze high-energy fast transients. Being the first GRB observed by EP, GRB 240219A illustrates the instrument's capability to investigate the soft X-ray component of GRB prompt emissions. Additionally, EP's recent detection of EP240617a \citep{2024GCN.36691....1Z}, in correlation with another untriggered XRR, GRB 240617A \citep{2024GCN.36692....1Y}, highlights its proficiency in identifying and researching XRRs. Together, these observations underline EP's substantial role in enhancing our understanding of GRBs with more attributes, particularly those that might otherwise remain unnoticed due to their weak gamma-ray signals or high redshift \citep{2024arXiv240416425L}.

\begin{acknowledgments}
We gratefully acknowledge the HXMT team for providing the Insight-HXMT light curve data of this event. We acknowledge the support by the National Key Research and Development Programs of China (2022YFF0711404, 2022SKA0130102, and 2021YFA0718500), the National SKA Program of China (2022SKA0130100), the National Natural Science Foundation of China (grant Nos. 11833003, U2038105, U1831135, 12121003, 12393811, and 13001106), the science research grants from the China Manned Space Project with NO. CMS-CSST-2021-B11, and the Fundamental Research Funds for the Central Universities. This work is supported by the CNSA program D050102. Part of the funding for GROND (both hardware as well as personnel) was generously granted from the Leibniz Prize to Prof. G. Hasinger (DFG grant HA 1850/28-1).
\end{acknowledgments}

\end{document}